**10**

# Florigens and antiflorigens: a molecular genetic understanding

## Ianis G. Matsoukas[1]

*School of Life Sciences, The University of Warwick, Gibbet Hill Campus, Coventry, CV47AL. United Kingdom.*

### Abstract

Florigens, the leaf-derived signals that initiate flowering, have been described as 'mysterious', 'elusive' and the 'Holy Grail' of plant biology. They are synthesized in response to appropriate photoperiods and move through the phloem tissue. It has been proposed that their composition is complex. The evidence that FLOWERING LOCUS T (FT) protein and its paralogue TWIN SISTER OF FT (TSF) act as florigen, or represent at least part of it, in diverse plant species has attracted considerable attention. In *Arabidopsis thaliana*, inductive photoperiodic conditions perceived in the leaf lead to stabilization of CONSTANS protein, which induces FT and TSF transcription. When they have been translated in the phloem companion cells, FT and TSF enter the phloem stream and are conveyed to the shoot apical meristem, where they act together with FLOWERING LOCUS D to activate transcription of floral meristem identity genes, resulting in floral initiation. At least part of this model is conserved, with some variations in several species. In addition to florigen(s), a systemic floral inhibitor or antiflorigen contributes to floral initiation. This chapter provides an overview of the different molecules that have been demonstrated to have florigenic or antiflorigenic functions in plants, and suggests possible directions for future research.

Q1

[1]*To whom correspondence should be addressed (email I.Matsoukas@bolton.ac.uk).*
*Systems and Synthetic Biology, Institute for Materials Research and Innovation, University of Bolton, Bolton, UK. Institute for Renewable Energy and Environmental Technologies, University of Bolton, Bolton, UK.*

**133**

# Introduction: the florigen mystery

The first historical record of attempts to understand flowering behaviour can be found in the writings of the Greek philosopher Theophrastus (c. 370–285 BC). Many centuries later, Mendel included flowering among the traits that he examined genetically, but his data were never presented in full. Detailed examination and quantification of flowering time variation began between 1910 and 1920, when the effects of low temperatures, nutritional conditions and daylength were first quantified.

The response of plants to the relative length of day and night is known as photoperiodism. Photoperiodism was first described in detail in 1920 by Garner and Allard, who demonstrated that plants flower in response to altered photoperiodic conditions [1]. The photoperiodic response allows plant species to adapt to seasonal changes in their environment [2]. Based on their flowering response, short-day plants (SDPs) flower after the daylength becomes shorter than a critical length (the inductive photoperiod), long-day plants (LDPs) flower when the daylength becomes longer than a critical length, and day-neutral plants flower irrespective of daylength. Both SDPs and LDPs can be either obligate (also termed qualitative) or facultative (also termed quantitative). Obligate plants have an absolute requirement for inductive photoperiods in order to flower, whereas the flowering of facultative plants is only accelerated in the inductive photoperiod [2]. *Arabidopsis thaliana* is a facultative LDP, as its flowering is promoted by long days (LDs) and delayed by short days (SDs).

Over the years, a number of studies have led to the development of three models to explain the photoperiodic regulation of floral induction. The first model is that of a flowering hormone-like substance, or florigen, which was first postulated by Mikhail Chailakhyan [3]. The florigen concept was based on the transmissibility of floral inductive signals across grafts between reproductive donor stems and juvenile recipients in *Nicotiana tabacum*. It was proposed that florigen was synthesized in the leaves under inductive photoperiods and transported to the shoot apical meristem (SAM) via the phloem. It was also proposed that florigen has a universal function in plants [4,5]. Many different molecules have been suggested to be components of florigen. The detection of a graft-transmissible floral antagonist also led to the theory of a competing 'antiflorigen.' Despite many years of research aimed at identifying the florigen and antiflorigen molecules in the phloem exudates of several plant species, their **Q2** molecular character has remained elusive until recently.

The difficulty in distinguishing the hypothetical floral hormone-like substance from phloem-transported assimilates led to the development of a second model, known as the nutrient diversion hypothesis. According to this model, floral-promotive conditions result in an increase in the amount of photosynthate that is translocated to the SAM, which in turn promotes floral induction.

The theory that photosynthate translocation is uniquely important in the promotion of floral induction was displaced by the multifactorial control hypothesis proposed by Georges Bernier [6,7]. According to this model, several promoters and/ or inhibitors belonging to the classes of nutrients and hormones are involved in floral induction in the SAM, and genetic variation as well as past and present environmental conditions result in different factor(s) becoming limiting in different genotypes and/or under diverse environmental conditions.

The molecular nature of florigen has long been a key research question in plant developmental biology. Interestingly, elegant molecular genetic approaches have recently pinpointed possible florigen and antiflorigen components. A vast body of literature suggests that FLOWERING LOCUS T (FT) protein [8–11] and its paralogue TWIN SISTER OF FT (TSF) [12] act as florigen, or at least that they represent part of it in a range of different plant species. The panoptic theme of floral induction pathways has attracted much attention, and a number of comprehensive review articles have been published [13–15]. However, this chapter will highlight the literature that has increased our understanding of several aspects of florigenic and antiflorigenic signalling in plants.

## Floral signal transduction

Flowering time has been genetically explored in several plant model systems, and many genes have been cloned through the study of natural variation and induced mutations (Figure 1). This has led to the conclusion that several interdependent genetic pathways control floral induction [13–15]. The photoperiodic and vernalization pathways control time to flowering in response to environmental signals such as daylength, light and temperature, whereas the autonomous and gibberellin (GA)-dependent pathways monitor endogenous indicators of the plant's age and physiological status. Other factors and less well-characterized pathways also play a role in the control of floral initiation. These include microRNAs (miRs), ethylene, brassinosteroids, salicylic acid (SA) and cytokinins (CKs). The photoperiodic pathway is of particular relevance to the focus of this chapter, and is therefore the only pathway that will be described here.                                                                      Q2

The actions of all flowering time pathways ultimately converge to control the expression of a small number of so-called floral pathway integrators (FPIs), which include FT, TSF, SUPPRESSOR OF CONSTANS 1 (SOC1) and AGAMOUS-LIKE 24 (AGL24). These act on the floral meristem identity (FMI) genes *LEAFY* (*LFY*), *FRUITFULL* (*FUL*) and *APETALA 1* (*AP1*), resulting in the initiation of flowering.

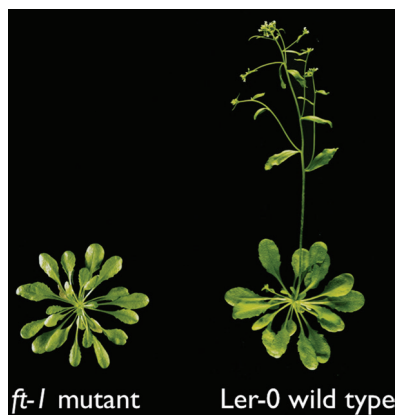

**Figure 1. The *ft-1* mutant of *Arabidopsis thaliana* flowers significantly later than the Ler-0 wild type under LD conditions.**

# The photoperiodic pathway: a model for the mode of action of FT

The photoperiodic pathway starts with the perception of light in leaves by the red/far-red light receptors, namely phytochromes (PHYA-E), and the blue/UV-A light receptors, namely cryptochromes (CRY1 and CRY2) [16]. The photoreceptors initiate signals that interact with a circadian clock and entrain the circadian rhythm. In plants, the circadian clock regulates a wide range of biological processes and represents the plant's endogenous timekeeper. Central to the photoperiod pathway in *Arabidopsis* is CONSTANS (CO), a zinc finger transcription factor (TF). Cycling of CO mRNA is regulated transcriptionally by the circadian clock through a protein complex formed by GIGANTEA (GI), and FLAVIN BINDING, KELCH REPEAT, F-BOX 1 (FKF1). The GI–FKF1 interaction modulates CO protein stability through degradation of the CO repressor CYCLING DOF FACTOR 1 (CDF1). High CO levels promote the expression of FT and TSF, two key members of the phosphatidylethanolamine-binding protein (PEBP) gene family [17–19].

Based on the temporal and spatial expression patterns of FT and TSF, a model for the photoperiodic induction of *Arabidopsis* has been proposed (Figure 2) [13–15]. According to this model, FT protein and TSF are part of the long-distance floral stimulus, florigen. Inductive LD conditions perceived in the leaf stabilize CO protein, which induces FT transcription in the leaf and TSF in the stem. Once they have been translated in the phloem companion cells, FT and TSF are loaded into the phloem and translocated to the SAM. FT-INTERACTING PROTEIN 1 (FTIP1) [20] and PHOSPHOLIPID PHOSPHATIDYLCHOLINE (PC) [21] have been identified as key regulators of FT transport. At the SAM, a series of direct interactions between FPIs, SQUAMOSA PROMOTER BINDING PROTEIN LIKE (SPL) TFs and FMI genes promotes the LD floral induction pathway [18,22–25]. FT and TSF have been shown to physically interact with the locally transcribed bZIP TF FLOWERING LOCUS D (FD). The FT–TSF/FD transcriptional complex activates the expression of *SOC1* [18]. After the induction of *SOC1*, expression of *SPL3, SPL4* and *SPL5* is rapidly induced in the SAM. These three members of the SPL family are direct targets of SOC1 and FD, whereas their expression also requires FT–TSF and SOC1–FUL activity [23,24]. When the transcription of FMI genes is sta-    **Q3** bilized, FT and TSF are no longer essential, and the SAM becomes fully committed to floral initiation. Under non-inductive SD conditions, FT expression levels are reduced. However, as plant development proceeds, FT expression levels show a clear increase.

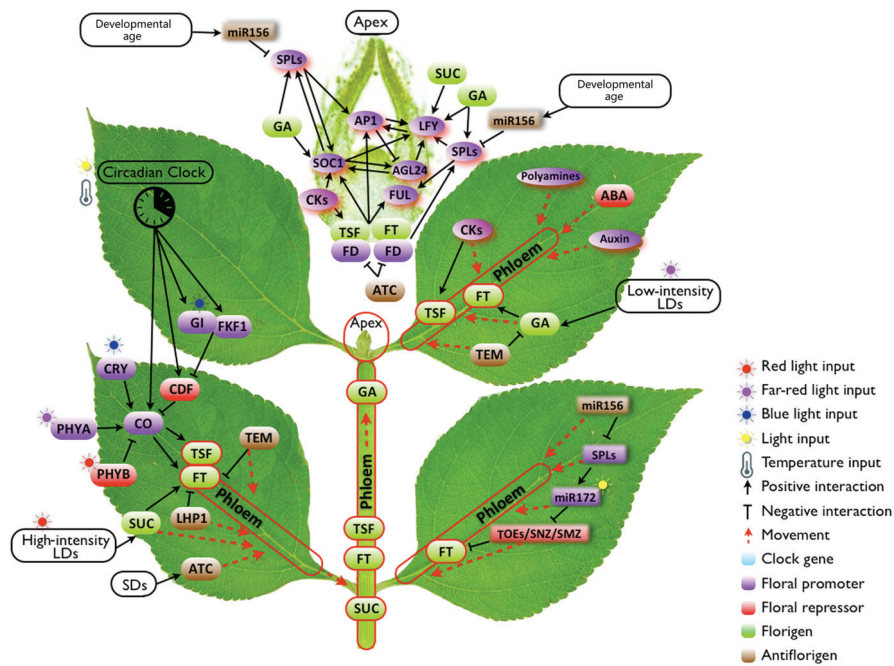

**Figure 2. Florigenic and antiflorigenic signal transduction pathways in *Arabidopsis*.**
Components of the signalling pathways are grouped into those that promote and those that repress the transition to flowering. The main florigenic and antiflorigenic components and interactions are depicted; additional elements have been omitted for clarity. Details are provided in the text.

Evidence has been provided for *FT* mRNA trafficking via the phloem to the SAM, independently of the FT protein [26]. Whether *FT* mRNA also participates in systemic floral regulation remains controversial [13]. However, under inductive LD conditions there is another florigenic signal involving specific GAs. Thus there may be multicomponent floral signalling in LDs involving FT and GAs, and an additional role for photosynthates has also been proposed (Figure 2) [13].

# The role of FT homologous, orthologous and paralogous genes in other species

At least part of the regulatory mechanism described above is conserved, with some variations, in several plant species. Loci orthologous to FT have been identified in several dicotyledonous and monocotyledonous plants. Species that possess *FT* genes include trees, woody perennials, grasses, legumes and ornamentals. Ectopic overexpression of *FT* orthologous genes hastens the juvenile-to-adult phase transition, and promotes time to flowering in several transgenic homologous and heterologous plants [13]. In addition, in several species the interaction between FT and FD has been either demonstrated or proposed [27,28].

Recent studies have revealed the persistently elusive florigens in species such as *Oryza sativa*, *Lycopersicon esculentum* and *Zea mays*. The *Z. mays CENTRORADIALIS 8* (*ZCN8*) gene is expressed in the leaf and is able to induce flowering in *Arabidopsis ft* mutants when expressed under the control of a phloem-specific promoter [29]. In *L. esculentum*, the *SINGLE FLOWER TRUSS* (*SFT*)-dependent graft-transmissible elements complement developmental defects in *sft* mutants, and substitute for LD conditions in *Arabidopsis* [8].

As an SDP, *O. sativa* requires different regulatory mechanisms for the photoperiodic control of flowering [30]. However, as in *Arabidopsis*, the central pathway in *O. sativa* consists of the *GI–CO–FT* regulatory module. The *O. sativa* homologues of *GI, CO* and *FT* are *OsGIGANTEA* (*OsGI*), *HEADING-DATE 1* (*HD1*) and *HEADING-DATE 3a* (*HD3a*), respectively (Figure 3) [31–34]. At the SAM, HD3a interacts with its intracellular receptor 14-3-3 protein, and binds to the OsFD1 to form a ternary 'florigen activation complex' [35], which induces transcription of FMI genes that lead to floral induction. *RICE FLOWERING LOCUS T*

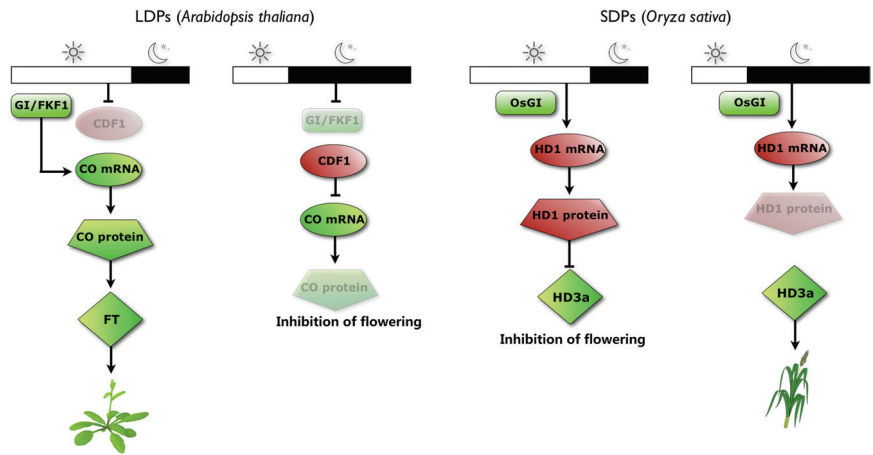

**Figure 3. A simplified model of the photoperiodic induction pathway in LD and SD plant species.**
Arrows indicate activation, and T-bars indicate inhibition. CDF1, CYCLING DOF FACTOR1; CO, CONSTANS; FKF1, FLAVIN-BINDING KELCH REPEAT F-BOX1; FT, FLOWERING LOCUS T; GI, GIGANTEA; HD1, HEADING-DATE1; HD3a, HEADING DATE 3a; OsGI, *Oryza sativa* GI. Details are provided in the text.

1 (*RFT1*), another orthologue of *Arabidopsis FT*, is a major floral promoter under LD conditions [36]. However, a unique *O. sativa* pathway with no obvious orthologue in *Arabidopsis* is also involved in photoperiodic regulation of floral induction. *EARLY HEADING DATE 1* (*EHD1*), a B-type response regulator, functions upstream of *HD3a* and *RFT1*. Mutations in *EHD1* cause delayed flowering under SD conditions [37]. In conclusion, this conservation, together with the small protein size, renders FT partially capable of fulfilling the requirements of one of the florigen candidates, or one of its important components.

# Florigen: beyond floral signal transduction

Photoperiod affects diverse developmental pathways in plants. A vast body of literature suggests that FT has universal effects on plant development. Recent studies have led to the identification of members of the FT gene family as a major component of the tuber-inducing signal, tuberigen [30]. Tuberization is a complex developmental process that initially leads to the formation of horizontally growing underground stems known as stolons. Tuberigen is first synthesized in the leaves under SD conditions, is graft-transmissible, and is transported through the phloem to the stolons. The molecular nature of tuberigen is not yet known, but it might be similar to that of florigen. Interestingly, *StSP6A*, a *Solanum tuberosum* FT homologue, has been shown to control tuberization [30]. Overexpression of the *O. sativa* florigen gene, *HD3a*, induced *StSP6A* expression and promoted tuberization as well as floral induction in *S. tuberosum* lines. This finding supports the hypothesis that florigen functions as tuberigen.

Florigen appears to play a broad pleiotropic role in the regulation of generalized growth of vegetative and reproductive organs. *FT* homologues are reportedly involved in SD-induced growth cessation and bud set in *Populus* species [19,38], in the control of leaf morphology and plant architecture in perennial *L. esculentum* [8] and *Z. mays* [39], in stomatal control in *Arabidopsis* [40], in fruit yield in *L. esculentum* [41], and in flowering repression in *Beta vulgaris* [42]. Interestingly, and perhaps understandably, the discovery of pleiotropic effects of FT genes has emerged as a field of research with significant potential to enhance several aspects of crop performance and quality.                                                                                    Q4

# Antiflorigenic signal transduction

The concept of certain antiflorigen(s) or floral repressor(s) was proposed almost as long ago as that of a floral stimulus [2]. Classical physiological studies have suggested the existence of certain antiflorigen(s) that are synthesized in leaves (Table 1). Interestingly, the PEBP gene family has evolved both activators and repressors of flowering.

Antiflorigenic effects of FT-like genes have been postulated in several plant species. *Arabidopsis thaliana RELATIVE OF CENTRORADIALIS* (*ATC*), an *Arabidopsis* FT paralogue, functions as an antiflorigen [28]. ATC probably antagonizes FT activity, because both ATC and FT interact with FD to regulate the same downstream FMI genes, but in an opposite manner [28]. Similarly, the SD plant *Chrysanthemum seticuspe* produces FT-like proteins with antagonistic functions. *C. seticuspe FT-like 3* (*CsFTL3*) is activated in SDs and promotes floral induction, whereas *C. seticuspe ANTI-FLORIGENIC FT* (*CsAFT*) is activated in non-inductive LDs and represses floral initiation [27]. The antagonism of CsAFT and CsFTL3 occurs through competition for CsFDL1, a *C. seticuspe* FD homologue. In *B. vulgaris*, time to flowering is regulated by the interplay of two paralogues of *Arabidopsis FT* that have evolved antagonistic functions. *B. vulgaris FT 2* (*BvFT2*), which is functionally conserved with *FT*, is essential for floral induction, whereas *BvFT1* acts as an antiflorigen [42]. Similarly, the *Helianthus annuus FT 1* (*HaFT1*) paralogue acts as a floral promoter, whereas the frame shift allele *HaFT4* functions as an antiflorigen [43].

Several studies have identified the antiflorigenic functions of LIKE HETEROCHROMATIN PROTEIN 1 (LHP1, also known as TERMINAL FLOWER2) [44], TEMPRANILLO 1 (TEM1) and TEM2 proteins [45]. It has been demonstrated that LHP1, TEM1 and TEM2 function as leaf-based antiflorigens that might also be able to move to the SAM. The LHP1 represses the expression of FT, but with no effect on the expression of TSF [12] or the other FPI and downstream FMI genes [44]. *TEM1* and *TEM2* genes play a key role in inhibiting flowering under SDs and LDs by directly repressing FT and GA biosynthesis genes [45].

miRs are short non-translated RNAs that are processed by Dicer-like proteins from large, characteristically folded precursor molecules. The majority of plant miRs target TFs, and therefore regulate several developmental events, including juvenility and floral induction [13,46]. Delicate grafting experiments have shown that the effect of several miRs is transmissible via grafts [47], which suggests that they are transportable. miR156 is an antiflorigen and key regulator of the juvenile-to-adult and vegetative-to-reproductive phase transitions in several plant species [48–50]. Constitutive expression of miR156 prolongs juvenility and time to flowering. It has been demonstrated that the juvenile-to-adult phase transition is accompanied by a decrease in miR156 levels, and a concomitant increase in the levels of miR172 and SPL TFs [51–54]. Expression of miR172 activates florigen in the leaves through repression of AP2-like TFs TARGET OF EAT1-3 (TOE1-3), SCHLAFMÜTZE (SMZ) and SCHNARCHZAPFEN (SNZ) [55–57], whereas the increase in *SPL* transcript levels in the SAM leads to the transcription of FMI genes [25,54]. Interestingly, miR156a and miR156c, which are the major sources of miR156 in *Arabidopsis*, are significantly down-regulated by sugars [58,59].

**Table 1. Summary of putative antiflorigenic signals**

| Gene name | Abbreviation | Species | Supporting evidence |
| --- | --- | --- | --- |
| *Arabidopsis thaliana* RELATIVE OF CENTRORADIALIS | *ATC* | *Arabidopsis thaliana* | ATC antagonizes FT activity. Both ATC and FT interact with FD |
| *Beta vulgaris FT2* | *BvFT1* | *Beta vulgaris FT1* | BvFT1 is likely to regulate downstream flowering promoters, as part of the vernalization response |
| *Chrysanthemum seticuspe ANTI-FLORIGENIC FT* | *CsAFT* | *Chrysanthemum seticuspe* | CsAFT antagonizes CsFTL3 activity. CsAFT and CsFTL3 compete for CsFDL1, an FD homologue |
| *Helianthus annuus FT4* | *HaFTA4* | *Helianthus annuus* | A potential mechanism for the negative interaction between the HaFT1 and HaFT4 could be interference with binding of HaFT4 to shoot apex proteins required for floral induction |
| MicroRNA156 | miRNA156 | *Arabidopsis thaliana, Oryza sativa, Zea mays* | In *Arabidopsis*, miR156 overexpression results in a prolonged juvenile phase and a delay in flowering. miR156 targets the transcripts of SQUAMOSA PROMOTER BINDING-LIKE (SPL) transcription factors |
| *TEMPRANILLO1/TEMPRANILLO2* | *TEM1/TEM2* | *Arabidopsis thaliana* | TEM1 and TEM2 repress flowering in SDs and LDs, by directly repressing FT and GA biosynthesis genes |
| *TERMINAL FLOWER2 (LIKE HETEROCHROMATIN PROTEIN1)* | *TFL2 (LHP1)* | *Arabidopsis thaliana* | LHP1 (TFL2) represses the expression of FT |

Antiflorigens are of great importance as they prolong the juvenile-to-adult and vegetative-to-reproductive phase transitions. This allows the necessary assimilate reserves to be accumulated, ensuring unimpeded reproductive development.

# Floral induction may involve multiple florigens and antiflorigens

The existence of long-distance signalling molecules for floral induction has been known for **Q5** several decades. Some of these molecules have florigenic or antiflorigenic functions, whereas others make only a floral-promoting or floral-repressing contribution (Table 2). Other compounds that have been postulated as graft-transmissible floral regulators are described below.

## Phytohormones in floral signal transduction

Among the phytohormones, GAs are of special importance because of their ability to induce flowering in LD plants grown under non-inductive SDs [60]. In *Lolium temulentum*, GA5 and GA6 have been demonstrated to be LD mobile floral signals that traffic to the SAM. GA4 promotes export of assimilates, and in combination with sucrose has a synergistic florigenic effect on the activation of FMI genes [61]. Moreover, GAs have important roles in promoting transcription of *FT, TSF* and *SPL* genes during floral induction in response to LDs. It has been shown that these functions are spatially separated between the leaf and the SAM [13].

CKs are another class of phytohormones with reported roles in regulation of flowering time in *Sinapis alba* and *Arabidopsis* [7,13]. Phloem sap analyses in *Arabidopsis* have revealed increased levels of the isopentenyl type of CKs during floral initiation in response to photoperiod. In *S. alba*, the florigenic effect of CK, which bypasses *Sinapis alba FT*, acts via its paralogue *TSF* and *SOC1* [62]. CKs and *SaFT* may therefore be integral parts of the florigen in *S. alba*.

Abscisic acid (ABA) was proposed as the first identified antiflorigen. *Arabidopsis* mutants that are defective in or insensitive to ABA flower early, whereas mutants that overproduce ABA flower late. The inhibitory effect of ABA on time to flowering might be explained by sugar repression-related events. This view is supported by the early flowering phenotype of ABA-deficient mutants and their allelism to sugar-insensitive mutants. However, despite its antiflorigenic function in some species, many other species are not affected, and therefore ABA seems to have no general function as a floral transmissible repressor [60].

Polyamines such as putrescine and spermidine might also represent part of the florigen. Photoperiodic induction of *S. alba* was correlated with a significant increase in the levels of putrescine, the major polyamine in the leaf phloem sap. Auxins have also been detected in phloem sap, and in conjunction with other molecules may have potential florigenic activity. The altered flowering phenotypes of plants with impaired auxin biosynthesis and signalling may indicate that modulation of flowering time by auxins is indirect, occurring via interaction with other phytohormones, such as CKs, ethylene, ABA and GAs [63]. Other molecules, such as ascorbic acid (AA), ethylene, salicylic acid (SA), brassinosteroids and peptides, have also been suggested to act as floral promoters [7,13,60].

**Table 2. Summary of putative components of florigen and antiflorigen**

| Molecule | Function | Supporting evidence | Conflicting evidence |
|---|---|---|---|
| Proteins | Florigenic | FT protein, as a florigen, shows universality across daylength response types | |
| | Differences in different proteins present in phloem exudates between induced and non-induced plants | | |
| | Transported from leaves to SAM | | |
| | Antiflorigenic | Repress flowering time | |
| | Differences in different proteins present in phloem exudates between induced and non-induced plants | | |
| | Transported from leaves to SAM | | |
| miRNAs | Florigenic | Promote flowering time (miRNA172) | |
| | Differences in different proteins present in phloem exudates between induced and non-induced plants | | |
| | Transported from leaves to SAM | | |
| | Antiflorigenic | Repress flowering time (miRNA156) | |
| | Differences in different proteins present in phloem exudates between induced and non-induced plants | | |
| | Transported from leaves to SAM | | |
| mRNAs (FT) | Floral promoting | Evidence shows that FT mRNA may act together with FT protein | Positive disproof of the mRNA hypothesis by using artificial miRNA against FT mRNA, and a synthetic FT gene with extensive nucleotide substitutions throughout the ORF and foreign UTRs, respectively |
| | FT mRNA promotes flowering in *Nicotiana tabacum* Maryland Mammoth. Transported from leaves to SAM. | | |
| Sucrose | Florigenic. Levels increase rapidly in leaf exudates under inductive conditions. Complements the late-flowering phenotype of *co* mutants. Transported from leaves to SAM. | Promotes flowering time in several plant species | Difficult to distinguish between signalling and metabolic effects |
| Peptides | Florigenic. Differences in peptides present in phloem exudates between induced and non-induced plants. Transported from leaves to SAM. Increase at SAM after increase in leaves and shortly, after transfer to inductive conditions. | Promote flowering time | The effects of most peptides on flowering have not yet been tested |
| Cytokinins | Florigenic. Detected in phloem sap. Levels increase at SAM shortly after induction. Transported from leaves to SAM. | Promote flowering time | Not sufficient to induce flowering when applied exogenously |
| Gibberellins | Florigenic. | Act as florigens in LD species | May not act as universal florigens across daylength response types |

*(Continued)*

**Table 2. Summary of putative components of florigen and antiflorigen** *(Continued)*

| Molecule | Function | Supporting evidence | Conflicting evidence |
|---|---|---|---|
| Ethylene | Transported from leaves to SAM.<br>Florigenic | Have a floral repressing role in some woody species of horticultural importance<br>Induces flowering in plant species belonging to the Bromeliaceae<br>It is not clear whether ethylene synthesis is induced under inductive photoperiods<br>Promote flowering time | Inhibits flowering in many other plant species |
| Brassinosteroids | Floral promoting<br>Detected in phloem sap. | | |
| Abscisic acid | Floral promoting<br><br>Floral repressing | Promotes flowering when given after or later in the dark period in *Pharbitis nil*<br>Inhibits flowering when given before or early in the dark period in *Pharbitis nil* | Inhibits flowering in many plant species<br>Promotes flowering in *Pharbitis nil* |
| Polyamines | Floral promoting | Putrecine levels are significantly increased in photoinduced leaves of *Sinapis alba* | Polyamines alone do not influence floral induction |
| Salicylic acid | Floral promoting<br><br><br>Detected in phloem sap. | Promotes flowering in *Arabidopsis* and several species of the Lemnaceae | Photoperiod-related changes in endogenous levels of salicylic acid have not been detected |
| Ascorbic acid | Floral promoting<br><br>Floral repressing | Low levels cause accelerated flowering under LDs<br>Low levels delay flowering in SDs through GA deficiency events | May not act as a universal floral promoter<br>May not act as a universal floral repressor |

## Sucrose as a florigen

Sucrose is the most extensively studied compound that may potentially participate in floral signal transduction [7,13,60]. It is the dominant transport metabolite for long-distance carbon transport between source and utilization sinks. In *Arabidopsis* and *S. alba* that are exposed to inductive LDs, sucrose levels increase rapidly and transiently in phloem leaf exudates. Defoliation experiments have demonstrated that the increase in sucrose export coincides with the start of mobile floral signal transport, and occurs before the activation of cell division at the SAM [64]. Ample evidence has been provided that sucrose may act dependently and/or independently of FT florigen [13,65]. *Arabidopsis* plants flower rapidly in SDs after exposure to 8–12 days at a high light integral. It has been shown that this 'photosynthetic' response is FT independent. In contrast, the *IDD8* locus of *Arabidopsis* was reported to have a role in FT-dependent induction of flowering by modulating sugar transport and metabolism via the regulation of *SUCROSE SYNTHASE4* activity [66].

Because sucrose affects many aspects of plant development, it is difficult to demonstrate whether its effects on floral signal transduction are direct or indirect. The broad-spectrum effect of sucrose is clouded by its dyadic function as a nutrient and signalling molecule, and by **Q6** the interaction between sucrose signalling and hormonal networks [60]. However, trehalose-6-phosphate (Tre6P), a metabolite of emerging significance, acts as a signal of sucrose status in plant tissues. Interestingly, *Arabidopsis* plants with impaired Tre6P signalling are late flowering. This late-flowering phenotype has been found to be caused by reduced expression levels of FT, elevated levels of miR156, and reduced levels of at least three miR156-regulated transcripts, namely SPL3, SPL4 and SPL5 [67].

# Conclusions

A major challenge in modern biotechnology is to develop new elite crop varieties with enhanced agronomic traits, such as optimal flowering timing and plant architecture to meet the growing demand for food, feed and biofuel resources. Decades of research on photoperiodic control of floral induction have greatly expanded our understanding of the molecular mechanisms that initiate and drive juvenility and floral induction in different species. It has been shown that juvenility and floral induction are controlled by long-distance communication via the phloem, which recruits a variety of florigenic and antiflorigenic signals. One of these signals, the FT protein, fulfils all of the criteria that were postulated for the flowering hormone florigen by Mikhail Chailakhyan. However, fundamental questions remain unanswered. The properties of florigen have been insufficiently studied, and the complete composition of florigen has not been established. Under inductive LD conditions, there is another florigenic signal that involves GAs. Thus, apart from the FT protein, there may be a multicomponent floral signal in LDs involving GAs and/or sugars. In addition, further research is needed to determine how the florigenic and antiflorigenic signals leave the phloem in order to reach their target tissues. It has been suggested that FT transport may also depend on other molecules, and that compounds other than FT and TSF might be components of florigen. **Q2**

FT and TSF are regulated by *CO*, are phloem mobile, and require other components to be present in the SAM to function as floral promoters. In addition, other proteins, as well as factors such as miRNAs, sugars, GAs and CKs, have roles in the activation of FMI genes that lead to flowering. Therefore the multifactorial florigen concept proposed by Georges Bernier [6,7] should be accommodated.

### Summary

- *Arabidopsis* FT protein is a component of florigen, a systemic signal that has been demonstrated to control photoperiodic flowering in diverse plant species.
- FT is synthesized in the leaves and selectively transported to the SAM. In the SAM cells, FT binds 14-3-3 and FD to form a ternary 'florigen activation complex', which activates FMI genes.
- Many different molecules have been postulated to be components of florigen, including sucrose, GAs, CKs, other plant hormones, certain amino acids, proteins, miRNAs and SA. Complex interactions between the signalling pathways that control the synthesis of these various florigenic molecules may occur.
- Photoperiodic regulation of floral initiation is also regulated by a systemic floral inhibitor or antiflorigen, which is exported from the non-induced leaves. The antiflorigen production system prevents precocious flowering.
- A more detailed understanding of the molecular mechanisms involved in florigenic and antiflorigenic signal transduction has many biotechnological applications.

Research in the laboratory of I.G. Matsoukas is supported by the University of Bolton, U.K.